An Assessment of "What does photon energy tell us about cellphone safety" by Dr. William Bruno

By Bernard Leikind

August 3, 2011

Published on EMF and Health:

http://www.emfandhealth.com/Do%20Cell%20Phones%20Cause%20Cancer.html

Also at: arXiv:1107.0086 [physics.bio-ph]

**Abstract:** Dr. William Bruno asserts the well-known fact that cell phones radiate microwaves in the classical regime. This, he says, means that the photon energy is not relevant to assessing safety. Citing optical tweezers as an example of biologically relevant non-thermal effects of electromagnetic radiation, Bruno concludes that all other reports of non-thermal effects from microwaves are likely valid. He seeks safety thresholds based upon requiring that cell phone energy density be less than $k_B T$. This proposal and related ideas produce thresholds many orders of magnitude below present values.

While Dr. Bruno is correct that cell phone microwave radiation is generally in the classical regime, he uses peculiar estimates (number of photons per cubic wavelength) that overstate the circumstance by more than 20 factors of ten. He misunderstands the operation of optical tweezers and ignores their significant thermal effects. He credulously accepts poorly supported claims of non-thermal effects. He mistakenly believes that $k_B T$ is the average thermal energy (per cubic wavelength or per cell) in materials. It is not. It is twice the average energy per molecule per degree of freedom in the material. The thermal energy density is $(1/2)k_B T$ X (average number of degrees of freedom of the molecules) X (Avogadro's number). (Avogadro's number is $6 \times 10^{26}$ molecules/mole.) Thus, Bruno's proposed safety thresholds are more than $10^{25}$ too low. Using the correct value for the average thermal energy would place the thresholds close to today's standards. Throughout his analysis he neglects the index of refraction of living tissue, $n \cong 9$, or the absorption length, $\alpha \cong 1$ cm$^{-1}$, in the microwave region.

Dr. William Bruno asks "What does photon energy tell us about cellphone safety?" in a pre-print submitted to the physics arXiv pre-print database on April 24, 2001.[1] Some in the scientific press noticed the paper and drew attention to it. Here is an example of friendly coverage from MIT's Technology Review, "Cell Phones, Microwaves and The Human Health Threat", on April 28, 2011.[2]

A pre-print is a research paper that scientists send to their friends and colleagues while simultaneously submitting the manuscript to a journal for publication. As the formal publication may take months, the pre-print is a way for a researcher to draw attention to his or her work rapidly. Unlike articles in the best scientific journals, professionals have not reviewed a pre-print for quality or importance. Therefore, it is rare for one to produce wide news coverage.

Coverage of a pre-print may be the result of the author's reputation as an important and distinguished scientist or may result from the importance of the results. Dr. Bruno describes himself as a physicist or as a theoretical biologist. He appears, however, to be nutty about the subject of electromagnetic fields and health effects. Santa Fe science writer George Johnson describes some of Bruno's activities, including a literal metallic hat for protection against the ever present electromagnetic fields of modern civilization, in this witty account from Slate, "On Top of Microwave Mountain: I tried to sauté my brain at the base of a cell phone tower. It didn't work."[3] The coverage is not the result of Bruno's scientific reputation.

The media interest arises from the significance of Bruno's claims: that there are important, so-called non-thermal effects that play a role when organisms absorb or interact with microwave radiation, and that scientists and regulators do not recognize these effects when they establish safety limits for cell phones, cell phone towers, or WiFi and other equipment. Hence, Bruno argues safety limits are much too high.

Physicists and other scientists have studied the interactions between electromagnetic fields and matter in all its forms from atoms and molecules to big chunks of stuff, pure and mixed up, for more than a hundred years. A major sub-field of this research deals with organisms and health effects. With the rise of cell phones, many people wonder if the phones' microwave radiation might have bad effects on our brains. People are particularly concerned about cancers. After all, the microwaves definitely penetrate the brain and, generally, people haven't been exposed to such radiation in their brains at any time in human existence. A vast industry has arisen in response to these concerns involving epidemiological research, bio-chemical and biological research, medical and clinical research, studies of rats and mice, studies of cells living in Petri dishes, and commercial enterprises to measure and protect us from the supposed dangers.

What do physicists know about this? Some forms of electromagnetic radiation are harmful and cause cancer. These are ultraviolet radiation, X-rays, and gamma rays. Ultraviolet radiation causes skin cancers. No other form of electromagnetic radiation causes any cancer. These other forms include visible light, infrared radiation (heat radiation, as some would describe it), microwaves, radio waves, and on to lower frequencies and longer wavelengths. While the biological scientists don't know everything about carcinogenesis, physicists know everything about how organisms absorb electromagnetic radiation and specifically microwaves.

Physicists point out that the forms of electromagnetic energy that cause cancers all have sufficient energy in their photons (like particles of light) that they can break chemical bonds and ionize atoms and molecules. No other forms of the radiation can do this. At frequencies below and wavelengths longer than the visible spectrum, physicists know that when any material (in the conditions of living organisms) absorbs electromagnetic radiation, the energy goes directly into the incessant and random jostling, vibrating, and twisting of the molecules. This is heating, or a thermal effect. There is no missing energy.

Since everyone knows that wearing a ski cap, swallowing a hot coffee, or jogging do not cause cancer, clearly cell phones, which produce less, sometimes much less, heating than these activities, cannot cause cancer. Some biological and epidemiological researchers, however, believe that they find deleterious effects of cell phone radiation. Therefore, they conclude that they must be seeing non-thermal effects.[4]

Sometimes these researchers are making an elementary blunder, assuming that when they do not detect a temperature increase in their experiments the effects must not be thermal in nature. It may be that they do not have sufficiently precise thermometers or do not measure the temperature in their experiments. Unfortunately, this error is common. Furthermore, when ice melts in a summer drink or water boils in a teakettle, the water's temperature doesn't change, yet these are both definitely thermal effects. A thermal effect may result in a detectable temperature increase, but it may not.

Other researchers seek non-thermal effects in otherwise well-known phenomena associated with terms such as resonance, multi-photon effects, or nonlinear effects. Many researchers have proposed such phenomena or effects, and this brings us to Bruno's paper.

Bruno points to the statements of some physicists that only the photons of ionizing radiation have sufficient energy to break chemical bonds. He says this is not the correct way to think about microwave radiation. While its photons are individually very weak, they are present in cell phone radiation in very large numbers. Using a peculiar measure of the photon population density, photons per cubic wavelength, he shows that X-rays and ultraviolet radiation from common sources (medical X-ray machines and the sun's radiation) have a relatively small number of photons per cubic wavelength, while microwave radiation from cell phones, cell phone towers have a relatively large number of them. This, he says, makes the individual photon energy less relevant than the tremendous numbers of photons flowing through tissues.

He chooses as an example of a non-thermal biological effect based upon the flow of large numbers of non-ionizing photons, the remarkable devices known as optical tweezers. These use lasers and backward microscopes to focus visible or infrared radiation to tiny spots. With careful optical design, the experimenters can use variations in the intensity of the radiation to move, pull, or twist tiny objects, even individual molecules such as proteins and DNA. The United States Energy Secretary Steven Chu won his Nobel laurels for his work with these devices. Intensity of radiation is another way of saying that there are large numbers of photons flowing.

Having established to his satisfaction that he has identified a non-thermal mechanism, because microwaves from a cell phone have many photons and optical tweezers involve many photons, Bruno then asserts that the existence of this mechanism shows that all of the researchers who claim to have found a non-thermal effect may be on to something. He lists putative breaches in the blood-brain barrier, microwave clicks, and several other effects.

Accepting these effects as real and as occurring well below the official microwave safety thresholds, Bruno says that safety limits based upon heating are inadequate. He seeks possible new principles on which to base the limits. He considers the natural microwave radiation from the sun, which is orders of magnitude below the radiation levels of cell phones. He considers what he says is the average thermal energy present in organisms, which he calculates is also orders of magnitude below cell phone thermal levels.

He sees no hope for safety in these thoughts, suggests future use of optical signals, and recommends hands-free use of cell phones as an otherwise inadequate move in the right direction.

Bruno begins his attempt to show that the photon energy is not, in itself, sufficient to judge the effects or safety of cell phones by estimating the significance of the quantum, particulate nature of the radiation as compared to certain other sources. The information in his Table 1 is here in my Table 1, which contains some clarifications and additions. His data is in *italics.* The symbol ~ means that the value is representative of a range.

Bruno demonstrates that there are many more photons per cubic wavelength in microwave radiation than in solar ultraviolet radiation at the earth's surface or in a typical X ray. He concludes that cell phone microwaves are in the *classical limit* while medical X rays and solar UV radiation are in the *quantum limit.* His second column shows that there is a difference of 44 orders of magnitude between the X rays and the microwaves! In this way of thinking, the chunkiness of the X rays, the quantum nature of electromagnetic radiation, is important for them, but the waviness of the microwaves, the classical nature of the electromagnetic radiation, is important for them.

No physicist would disagree that in certain contexts X-rays and UV tend to exhibit quantum effects, chunkiness, or that the microwaves tend to exhibit classical effects, waviness. But Bruno mistakes these tendencies with invariable behavior. Scientists use X ray diffraction, which is a classical wave effect as an important probe of crystal structure. Microwave engineers don't worry about microwave photons for their purposes, but the photons are there nonetheless, and physicists studying the energy states of rotating molecules must consider microwave photons.

In his desire to illustrate that microwaves are far from the quantum realm of X-rays and ultraviolet, Bruno uses a peculiar measure of the photon density, photons per cubic wavelength. (See the note at the end to convert between Bruno's units and ones that are more usual.) Photons are an odd concept in quantum physics. It is sometimes allowable to imagine them as tiny little balls or perhaps even points. The famous Heisenberg Uncertainty Principle, the enforcer of quantum fuzziness, however, makes it hard to say just where a photon is in volumes smaller than a cubic wavelength. There is, however, no problem with estimating how many of a particular type of photon there might be in larger volumes. Bruno wishes to show that there are many, many more microwave photons around than X-ray photons. He counts how many microwave photons there are in a cubic microwave wavelength, which is, for his estimate, 0.3 meters, and cubed is 0.027 m$^3$ = 2.7 X 10$^{-2}$ m$^3$. He also counts how many X-ray photons there are in a cubic X ray wavelength, which is, for his estimate, 10$^{-10}$ m, and cubed is 10$^{-30}$ m$^3$. He is mistaking the size of a photon for how accurately you can say where it is, and by this mistake, he gets a factor of $10^{28}$ increase in the number of microwave photons relative to the number of X ray photons.

In Table 1 I have shown the Wavelength Cubed in parentheses in the second column so that you can see this factor.

Any physicist seeking clarity in comparing the photon density or number of photons would have chosen the same volume for each case. I have added column 3 to show this quantity, the photon density in photons/m$^3$. The physicist would also choose the same power or energy. I have added column 4, Photons/sec in 1 Watt = 1 Joule/sec. I can't tell what X ray power level Bruno had in mind, but I believe that 1 Watt of X-ray power is in the ballpark for a typical medical exposure. Bruno says that solar ultraviolet radiation is about 10 W/m$^2$, a flux or flow. (See the note at the end to convert between Bruno's power flow, Watts/m$^2$, and energy density, Joules/m$^3$.) This is also in the ballpark for a 1 Watt nearby source. Cell towers broadcast more than 1 Watt, but not much more, and all exposures are from many meters away. Bruno takes 10 meters, as if a person were standing at the base of the tower. Cell phones broadcast about one Watt when they are transmitting. The fourth column shows the number of photons for each of the exemplary sources for the same emitted power. It is the case that microwaves have about ten orders of magnitude more photons than the same X ray power, but ten orders of magnitude are 34 orders of magnitude smaller than the exaggerated 44 orders Bruno gets from his volume factor. I have also added a row that describes optical tweezers, since Bruno chose them as an example of a non-thermal source of electromagnetic radiation.

While objecting to Bruno's calculation, accept, for the moment, his point that microwaves are usually in the classical, wave-like limit. He says that optical tweezers are an example of a optical system that operates in the classical limit and that is capable of exerting forces on molecules that might well damage them without heating; a non-thermal effect.

**Table 1***

| Source | Photons per cubic wavelength (Wavelength Cubed m$^3$) | Photon Density Photons/m$^3$ | Photons/second in 1 Watt | Notes |
|---|---|---|---|---|
| *Medical X-ray* | *~ 1 X 10$^{-24}$* (10$^{-30}$) | 10$^6$ | 1.5 X 10$^{15}$ | *X-ray: 30 cm from 1mA source, 1% efficient* *Wavelength = 1 nm = 10$^{-10}$ m* |
| *Sunlight UV* | *~ 1 X 10$^{-7}$* (2.7 X 10$^{-24}$) | 4 X 10$^{14}$ | 1.5 X 10$^{17}$ | *~10W/m$^2$* *Wavelength = 300nm = 3 X 10$^{-7}$ m* |
| Optical Tweezers (Visible or Infrared) | *~ 1 X 10$^{-1}$* (1 X 10$^{-18}$) | 10$^{17}$ | 1.5 X 10$^{18}$ | 1 W visible or infrared laser beam focused to a 3 X 10$^{-6}$ m radius beam. Wavelength = 3000 nm = 3 X 10$^{-6}$ m |
| *Cell tower* | *~ 1 X 10$^{+15}$* (2.7 X 10$^{-2}$) | 4 X 10$^{16}$ | 1.5 X 10$^{24}$ | *Frequency = 1 GHz* *Wavelength = 3 X 10$^{-1}$ m* *E = ~1V/m* *10 meters from the antenna* or about 33 feet |
| *Cell phone* | *~ 1 X 10$^{+20}$* (2.7 X 10$^{-2}$) | 4 X 10$^{21}$ | 1.5 X 10$^{24}$ | *Frequency = 1 GHz* *Wavelength = 3 X 10$^{-1}$ m* *E = ~300V/m* (10 m / 300) = 0.03 m or about 1 inch from the antenna |

*Items, data and notes from Bruno's Table 1 are in *italics.* I added other items, data and notes.

Figure 1 a) shows a simplified schematic of an optical tweezers system, a laser with its beam passing backwards through a microscope to focus to a tiny spot. Figure 1 b) shows the details of the radiation

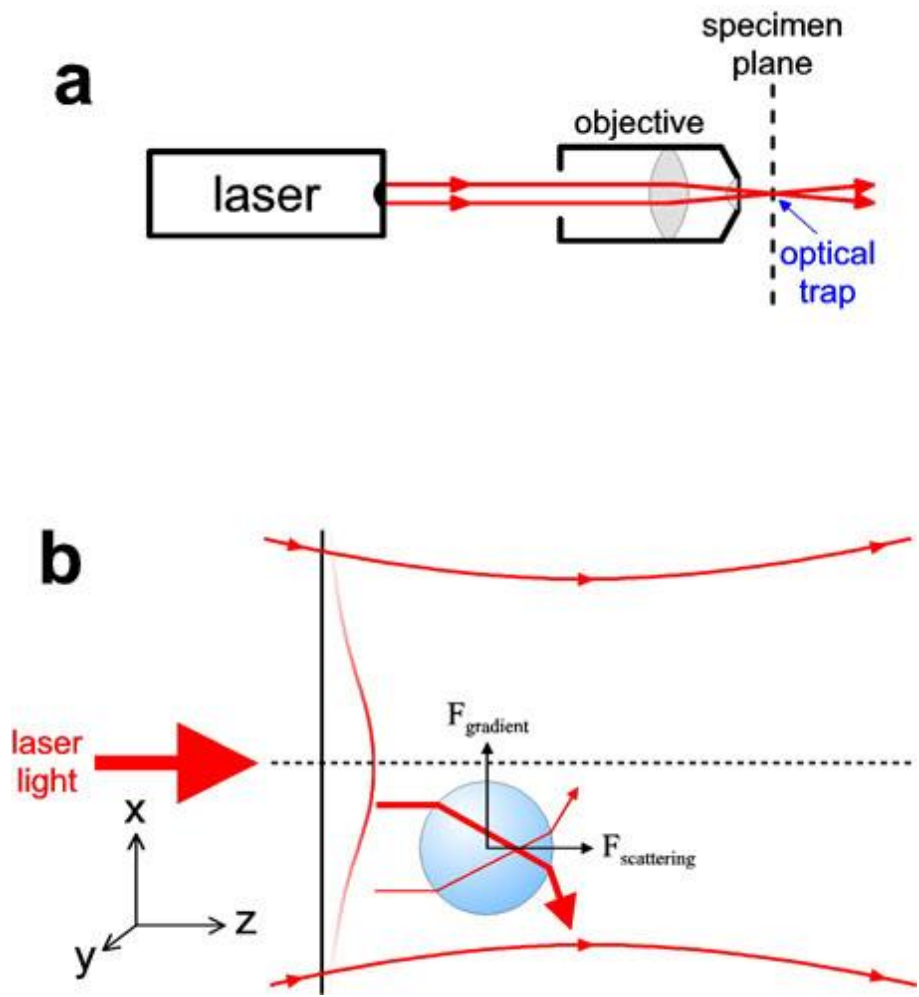

Figure 1. Optical Tweezers principles.

**How does it work?** The most basic form of an optical trap is diagramed in Fig 1a. A laser beam is focused by a high-quality microscope objective to a spot in the specimen plane. This spot creates an "optical trap" which is able to hold a small particle at its center. The forces felt by this particle consist of the light scattering and gradient forces due to the interaction of the particle with the light (Fig 1b, see Details). Most frequently, optical tweezers are built by modifying a standard optical microscope. These instruments have evolved from simple tools to manipulate micron-sized objects to sophisticated devices under computer-control that can measure displacements and forces with high precision and accuracy.

Figure 1. An illustration of the operation of an optical tweezers system and some explanatory text.[5]

beam in the spot or focused neck and a spherical object on which the beam exerts a force drawing the object to the center of the beam and to the narrowest part of the beam. Further details and a clear explanation of the operation of optical tweezers and of some of their applications are at the Stanford University site[5].

Considering round numbers, a typical optical tweezer system will use a 1 watt laser radiating in the visible or infrared region. It will pass the laser's beam backwards through a microscope to shrink it to a small region, a few microns, millionths of a meter, in diameter. Let's consider 6 microns in diameter, divide by two to get the radius, square it, and multiply by pi, to get the cross section of the beam. It is about $30 \times 10^{-12}$ m$^2$. Now compute the power flow (1 watt)/( $30 \times 10^{-12}$ m$^2$ ) = $3 \times 10^{10}$ W/m$^2$. This is 30 billion watts per square meter, a very intense beam. Bruno says that a medical X-ray has about 10 watts per square meter. The total power radiated from a cell phone is about one watt. Consider, just to estimate, that the cell phone's antenna is a small source, and imagine a sphere 1 cm = 0.01 m in radius. The surface area of this sphere is (4/3) pi r$^2$ = $4 \times 10^{-4}$ m$^2$. Now compute the power flow for this case (1 watt)/( $4 \times 10^{-4}$ m$^2$ ) = $0.25 \times 10^{4}$ w/m$^2$ or about 250 W/m$^2$. Two hundred and fifty is smaller than 30 billion by about a hundred million. The flow of photons per square meter radiated from a cell phone is smaller than the flow of photons per square meter in the tweezers' focal spot by the same ratio.

The effects of the optical tweezers, the ability of this intense beam of visible or infrared light, to exert forces on tiny objects depend upon spatial variations in the intensity of the tiny beam and on variations in the optical properties of the illuminated material. The intensity is highest along the axis of the optical system. In Figure 1 b) you can see the variation in the beam's intensity as the curved mostly vertical line to the right of the big red arrow. Imagine the red lines rotated about the dotted line to picture the generally cylindrical form of the optical beam.

The relevant optical properties of a material have two parts, one called the index of refraction, which is the part that matters for the laser to exert its force. This force would be the Bruno's non-thermal effect if it were the only relevant effect. The other part of the transparency is the absorption coefficient. This coefficient measures how far a beam of radiation travels before about 2/3rds of it has been absorbed. Each time the beam travels a distance equal to the absorption coefficient, its intensity falls another 2/3rds. The energy that disappears from the beam appears in the material as thermal energy.

The optical tweezers system depends upon the difference in the index of refraction in the object (the blue sphere shown as a circle in the figure) and in the background material. You can see that the variation in the intensity of the laser beam and the variation in the index or refraction, the size of the object, are roughly, the same scale.

In the visible and infrared range chosen for optical tweezers, the transparent materials do not absorb much of the beam, but they absorb some. A beam might travel many meters before a noticeable amount is lost. The radiation flow is so intense, however, that researchers risk damaging the materials, molecules or organisms they study because of the heating. That is, Bruno's example of a non-thermal effect often comes with important thermal side effects.

The situation is different in the case of microwaves and the tissues of living organisms. Water molecules love microwaves, as do many other biologically important molecules. The absorption coefficient is much higher, a few hundred times higher, than it is for visible or infrared light. Therefore, the distance the radiation travels before the materials absorb it is much shorter. Nearly all of the radiation from a cell phone that enters an organism disappears in a few centimeters. All of this is a thermal effect. There is no non-thermal effect. All of the radiation moves from the microwave beam into the thermal motions of all of the molecules of the tissues.

Having incorrectly convinced himself that optical tweezers provide an example of a relevant non-thermal effect, Bruno accepts, therefore, all other reports of non-thermal effects and concludes that experts setting radiation safety thresholds must consider these reported effects. He turns to investigating possible new safety thresholds, well below those set with regard to thermal effects. In this arena, he also chooses odd contexts and in one case makes an outright and major blunder.

He considers natural microwave fluxes, such as those present from the sun, he says. But he notes that these levels would be even lower at night. He doesn't think it would be practical to limit cell phone radiation to these levels, but he gives no detail.

He considers the "average thermal energy, $k_b T$, per cubic wavelength" in the organism. As long as the average energy from the cell phone is less than this quantity, he says, the organism should be safe. In this quantity $k_b$ is the Boltzmann constant, about $1.4 \times 10^{-23}$ Joules/K, and T is the temperature measured in Kelvins. Body temperature is about 310 K. This is an amazing mistake. The average thermal energy density is not $k_b T$. That value, $k_b T$, is twice the energy per degree of freedom of a single molecule. Bruno should have multiplied this by the number of molecules in the volume he's interested in and by the average number of degrees of freedom for the molecules in his volume. This latter is some small integer, let's just say 5, which would be appropriate for a water molecule. Remember, too, that a cubic wavelength for Bruno's cell phone (see Table 1) is about 27 liters (a cube 30 cm or a foot on a side).

The immense error is neglecting the number of molecules in his volume. Just to get a value, let's suppose a brain has the density of water, which would not be far off. A brain has a mass of about 1.2 kg and a volume of 1200 $cm^3$. This is just about a cubic wavelength for a 12 cm 2.5 GHz microwave. Water has about $3 \times 10^{22}$ molecules per gm or per $cm^3$. In a cubic wavelength there would be about $4 \times 10^{25}$ molecules. Bruno's estimate for a threshold safe limit is too low by a factor about 40 million billion billion! (That's US billions, 1,000,000,000.) That's wrong by 40 septillions. Don't forget that there is another factor of 10 or so for the degrees of freedom.

If Bruno were to use the correct value here, he'd come up with something similar to modern day safety thresholds. Experts set those standards by requiring that the energy or temperature changes in organisms caused by microwaves should be less than normal variations that result from ordinary activities.

Bruno cites various studies that claim to have seen effects, some deleterious, even at his wildly mistaken values. He cites reports of vague symptoms, headaches, depression, and sleeplessness, from people within a few miles of cell towers. These people, he says, experience microwave levels similar to his mistaken threshold. This tells us that those studies are foolishly in error, and the reports of symptoms are mistakenly attributing the symptoms to the cell tower radiation. We know that these studies and reports are foolish and mistaken because the reported effects are the result of radiation energy density many orders of magnitude less than what is naturally present within living organisms.

In his entire consideration of microwaves propagating in living tissue, Bruno has neglected a major effect that I have also neglected so far because it would be too much to correct every error at once. I have discussed the value of the absorption coefficient but I have not mentioned the value of the index of refraction for microwaves in living tissue. Of course, in this case, we are considering a complex material that includes skin, bone, brain, and so on, and the details would be complicated. The matter is further complicated because the absorbing material is within a few wavelengths of the antenna. But the value of the index of refraction that Bruno should be using in his analysis is about nine.[6] This means that in the tissue, the wavelength will be, roughly speaking, about nine times shorter than it is in air. He has been over estimating the volume of a cubic wavelength, in an organism, by about 9 X 9 X 9 or about 700. Astonishingly, this error is so many orders of magnitude less than his other errors and miscues that neglecting it hasn't changed my conclusions.

Finally, Bruno points to the possibility of damaging effects arising from the modulation of the cell phone microwaves. This suggests that he, and the researchers he cites, have no idea about the modulation of cell phone signals. Modern digital cell phone signals use complex modulation schemes, but they are not turned on for a one and off for a zero. They shift between two frequencies that barely differ. The frequency might be a bit high for a one and a bit low for a zero. The main carrier frequency is on all the time that a phone call is in progress. In actual use, there are various nearby, in frequency, channels. A call may switch among channels during the call, and so on. But there is always a high frequency carrier signal. The modulation certainly makes no difference whatsoever in the absorption of a cell phone signal.

**Note about units and quantities**

To convert from Bruno's photon number density, in photons per cubic wavelength, to a more usual photon number density, $n$ = photons per cubic meter, multiply his quantity by the cube of the number of wavelengths in one meter. The number of wavelengths in one meter is one divided by the wavelength. Thus

*Photon density* = # / wavelengths$^3$ [photons / cubic wavelength] X $(1/\lambda^3)$ [wavelengths / meter]$^3$ = $n$ [photons/ meter$^3$],

where # represents the number of photons in the volume, $\lambda$ is the wavelength, n is the photon density in #/m$^3$, and I have put the units in brackets [] and multiplied out the units as if they were algebraic quantities.

To convert from a density, *n*, to a flow, *j*, multiply the density by the speed of the flow, *nv = j.* In the case of electromagnetic radiation, the speed is *c*, the speed of light, $3 \times 10^8$ m/s. For example, to convert an energy density, in joules / $m^3$, to an energy flow or power flow, in Joules/s/$m^2$ or Watts/$m^2$, where a Watt is a Joule/s, multiply *n* [Joules/$m^3$] X $3 \times 10^8$ [m/s] = ($3 \times 10^8$) (*n*) [Joules/s/$m^2$ = Watts/$m^2$]. In this case, I've shown the conversion for an energy density to an energy (or power) flow. The conversion will work also for a photon density to a photon flow or flux, which would convert [photons/$m^3$] to [photons/s/$m^2$].

**References**

1. Bruno, William. "What does photon energy tell us about cellphone safety?", April 24, 2001. arXiv:1104.5008v1 [q-bio.OT].
2. KFC. "Cell Phones, Microwaves And The Human Health Threat", Physics arXiv blog at MIT's Technology Review, April 28, 2011. http://www.technologyreview.biz/blog/arxiv/26708/ accessed June 30, 2011.
3. Johnson, George. "On Top of Microwave Mountain: I tried to sauté my brain at the base of a cell phone tower. It didn't work."[3] Slate, April 21, 2010, http://www.slate.com/id/2251432/ accessed June 30, 2001.
4. Leikind, Bernard. "Do Cell Phones Cause Cancer?", Skeptic Vol. 15, No. 4, 2010, p.30. Leikind, Bernard. "Do Cell Phones Cause Cancer?", eSkeptic, June 9, 2010, arXiv:1007:4192 . http://www.skeptic.com/eskeptic/10-06-09/ , Shermer, Michael and Bernard Leikind, "Cell Phones and Cancer", eSkeptic http://www.skeptic.com/eskeptic/10-12-08/ , both accessed June 30, 2011.
5. From http://www.stanford.edu/group/blocklab/Optical%20Tweezers%20Introduction.htm , accessed June27, 2011. Used by permission.
6. See Jackson, J. D. *Classical Electrodynamics,* 2nd ed., John Wiley & Sons, Inc., New York, 1975. Pages 290-2, which discuss the index of refraction and absorption length for water, including sea water, which will serve as a useful approximation to the value for living tissue.